# Two-phase interfacial structure of bubbly-to-slug transition flows in a 12.7 mm ID vertical tube

Zhuoran Dang[a,*], Guanyi Wang[a], Mamoru Ishii[a,*]

[a]*School of Nuclear Engineering, Purdue University, 516 Northwestern Ave., West Lafayette, IN, 47907, USA*

**Abstract**

This experimental study focuses on the characteristics of air-water two-phase interfacial structure of bubbly-to-slug transition flow. Interfacial parameters including void fraction, interfacial area concentration, and bubble interfacial velocity are measured using four-sensor electrical conductivity probe on a 12.7 mm ID vertical tube. The tube size is approximately equal to the maximum distorted bubble size. Therefore, the bubbly-to-slug transition characteristics can be unique from in other sizes of tubes. Comparing with previous studies, this study provides an experimental database with a wide range of the bubbly-to-slug transition regime, with 4 different superficial liquid velocities (0.3, 0.5, 1.0, and 2.0 *m/s*) and void fraction ranging from 0.07 to 0.66. Experimental results show that the wall-peak void distribution does not appear in a small diameter tube under the bubbly-to-slug flow transition flow. The distribution is related to both void fraction and the relative bubble size to the tube size. In this sense, a new correlation of distribution parameter in the Drift Flux model is proposed based on the previous studies by Ishii, and Hibiki et al. This experimental study can be a good reference for the model development of flow regime transition and the Interfacial Area Transport Equation.

[*]Corresponding authors

*Email addresses:* zdang@purdue.edu (Zhuoran Dang), ishii@purdue.edu (Mamoru Ishii)

**Nomenclature**

$a_i$     interfacial area concentration

$D_c$     maximum distorted bubble size

$D_h$     channel hydraulic diameter

$D_{sm}$     Bubble Sauter mean diameter

$g$     gravitational acceleration

$j$     superficial velocity

$R$     tube radius

$r$     radial position

$v_i$     interfacial velocity

**Greek Symbols**

$\alpha$     void fraction

$\rho$     density

$\sigma$     surface tension

**Subscripts**

1     Group-1

2     Group-2

$f$     liquid phase

$g$     gas phase

# 1. Introduction

Two-phase flow regime transition is a complex physical process. It happens when the two-phase interfacial structure changes, which is related to many thermal-hydraulic properties including phase fraction, flow rate, pressure, and heat transfer rate, etc. An accurate prediction of the flow regime transition is of vital importance to the establishment of two-phase flow models. Currently, there is no universal theory on the flow regime transition. Many studies have been performed in terms of the transition between different flow regimes and a lot of empirical and semi-empirical methods for determining the flow regime transition have been developed. [1, 2, 3, 4] The application ranges of these methods are limited by the flow conditions and flow channel geometries used in their studies. In this sense, the development of flow regime transition models requires experimental databases with various flow conditions and flow channel geometries.

Another approach that models the flow regime transition is to use the Interfacial Area Transport Equation (IATE).[5] The biggest difference between the IATE and the flow regime transition criteria is that the IATE uses the bubble interaction mechanisms to dynamically predict the interfacial behaviors. It has been well proved that the IATE can give good predictions for bubbly and bubbly-to-cap-bubbly flows in vertical small tubes with the tube diameters between 25.4 to 50.8 mm.[6, 7, 8, 9, 10] However, the IATE still has many potentials for bubbly-to-slug transition flows. To fully realize the advantage of the IATE on dynamically predicting two-phase flow, more analyses are needed that require more rigorous experimental data with various flow conditions and channel geometries as support.

In the past experimental studies on the characteristics of the interfacial structure in vertical round tube, there is no available experimental data or discussions of the flow regime transitions in a 12.7 mm ID tube.[11] However, this size of the flow channel can be important in terms of bubbly-to-slug flow regime transition. Under the adiabatic, air-water flow condition, the hydraulic diameter of the flow channel is approximately equal to the maximum distorted bubble diameter, $D_h \approx D_c$. $D_c$ is proposed by Ishii and Zuber [12] and defined as

$$D_c = 4\sqrt{\frac{\sigma}{g\Delta\rho}} \quad (1)$$

For the air-water flow at atmospheric pressure and 20 °C, the maximum distorted bubble diameter is 10.9 mm from Eq. 1. This special flow channel geometry can give unique flow characteristics compared with other sizes of flow channels. For example, when bubble diameter reaches to $D_c$ due to the bubble coalescence and expansion, bubbles are constrained by the wall and directly transfer into slug bubbles. Thus, there can be no cap bubble in this size of flow channel theoretically. In this study, the bubbly to-slug transition characteristics are analyzed in a 12.7 mm ID tube. The motivation of this study is to use a broad range of experimental data collected and characterize the bubbly-to-slug transition behaviors under different void fractions, superficial liquid, and gas velocities. The collected data can be used for the development of the IATE model, as well as two-phase flow regime transition models in small diameter flow channels.

## 2. Experimental

### 2.1. Experimental facility

The test facility in the present study is designed and modified based on previous work [13]. The schematic diagram of the experimental facility is shown in Fig. 1. The main components include a centrifugal pump, a two-phase mixture injection unit, a test section, a water tank, a compressed air system, a piping system, measurement instruments, and a data acquisition system.

The water flow is supplied by a 25 hp centrifugal pump, the frequency of which can be preciously controlled. The water flow rate is measured by an electromagnetic flowmeter, with a measurement uncertainty of ±1%. The air is supplied by an external air compressor and the air flow rate is measured by 4 rotameters with different ranges, covering the superficial air flow rate up to 27 m/s. The measurement uncertainty of each rotameter is ±2% of full scale. The injection system is used for mixing water and air and injecting the mixture into the test section. A conceptual diagram of the injection unit is given in Fig. 2. The water pumped from the water tank passes the buffer tank and is separated into the primary and secondary water flow. The purpose of secondary flow is to shear the air that flows through the porous media. The sheared-off air is in the form of small spherical bubbles and its diameter is related to the secondary flow rate. In this experiment, the secondary flow rate is maintained at around 0.1 m/s and the sheared-off bubble diameter is around 1-2 mm. [14] The primary water flow is injected into the injector at the upper level and mixes with the two-phase flow before the test section inlet. The test section is made of an acrylic tube with an inner diameter of 12.7 mm. Local two-phase flow measurements are performed using four-sensor conductivity probes at each of the three axial locations $z/D_h =$ 31, 156, 282, and 10 radial positions from $r/R = 0$ to 0.875. Each measurement location also contains an impedance meter for measuring area-averaged void fraction and a pressure tap for measuring local pressure. The local pressure is measured using differential pressure (DP) gauges (Honeywell). The pressure drops along the test section are obtained from the pressure differences between local pressures and the pressure at the inlet. The measurement uncertainty is ±0.025% of the total setup range. After exiting the test section, the two-phase mixture goes into the storage tank, where water and air are separated and the fluid cycle is finished. The experimental data is collected using the data acquisition boards (NI USB-6255, National Instruments, Austin, TX) with a measurement frequency of 50 kHz and a data recording time of 60 seconds. These boards are connected with a computer for data reading and recording.

Four-sensor conductivity probe [15, 16] is used for the measurement of two-phase flow parameters, including time-averaged void fraction, interfacial area concentration, and bubble interfacial velocity. The schematic of the four-sensor conductivity probe used in the current study is given in Fig. 3. The principle of the four-sensor conductivity probe in measuring the void fraction is to calculate the ratio of the bubble residence time measured by the leading sensor to the time interval considered. [17] The principle of measuring the bubble interfacial velocity can be simply considered to be the time needed for the interface passing sensor tips. A detailed discussion of the interfacial velocity measurement principle can be found in Shen et al. [18]. The principle of the four-sensor probe in measuring the interfacial area concentration was first proposed by Kataoka et al. [15]. Later on, Revankar and Ishii [19], Kim et al. [16], and Shen et al. [18] further improved this method. Previous study on benchmarking the four-sensor conductivity probe in measuring the IAC shows that the uncertainty is less than ±10%.[16] Following the previous study [13], the bubbles measured by the four-sensor probe are classified into two groups using the maximum distorted bubble diameter $D_c$, expressed in Eq. 1: Group-1 bubbles are small spherical bubbles, and Group-2 bubbles are large distorted bubbles. The difference in bubble size and shape leads to substantial differences in transport phenomena due to the differences in drag force and liquid-bubble interaction mechanisms [20].

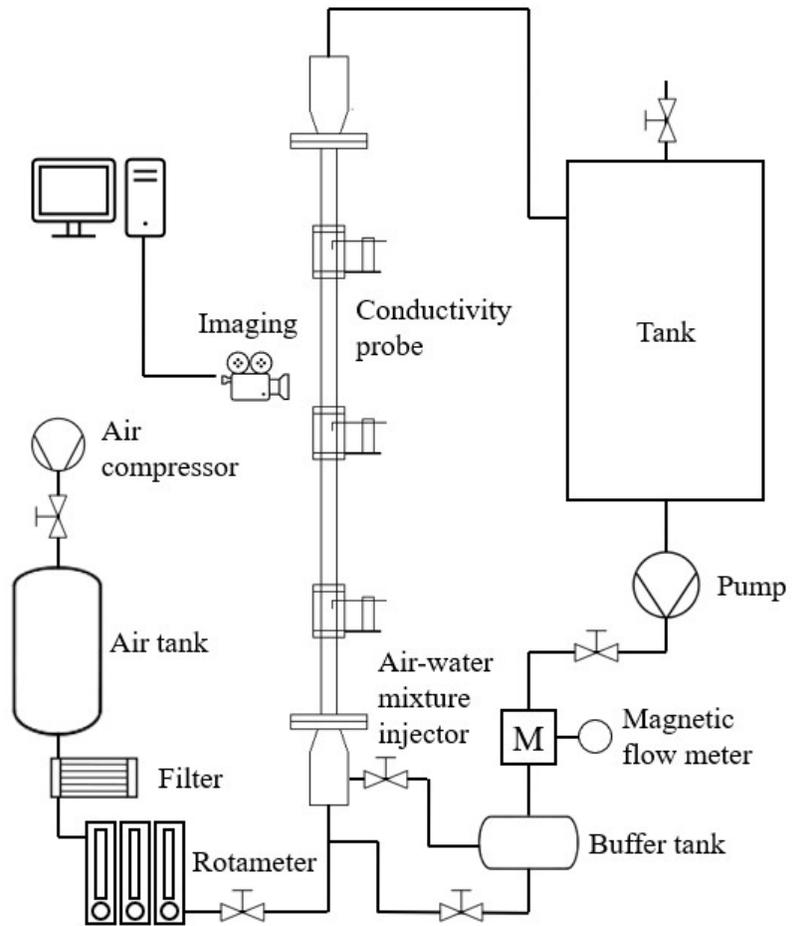

Figure 1: Schematic of experimental facility.

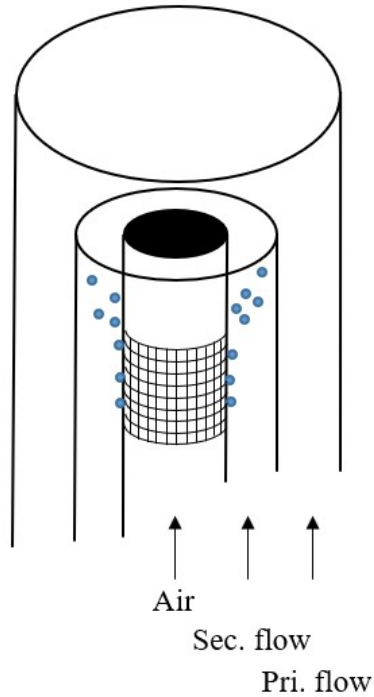

Figure 2: Conceptual schematic of the air-water mixture injector.

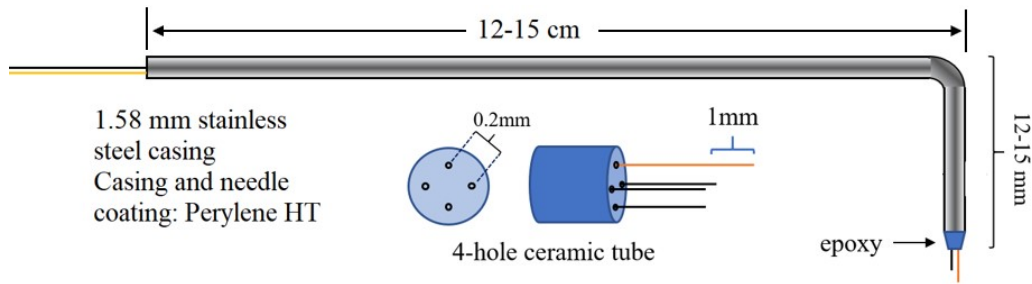

Figure 3: Four-sensor electrical conductivity probe.

## 2.2. Test matrix

The flow conditions are selected based on the Mishima-Ishii flow regime map.[2] 23 flow conditions in total are performed in this experiment and they belong to bubbly-to-slug flow transition regime, as shown in Fig. 4. The void fraction contour lines ($\alpha = 0.15, 0.2, 0.25$) are also plotted on the flow regime map as references. These lines are obtained based on the Drift-Flux model [21]. Detailed information about these flow conditions are given in Table 1.

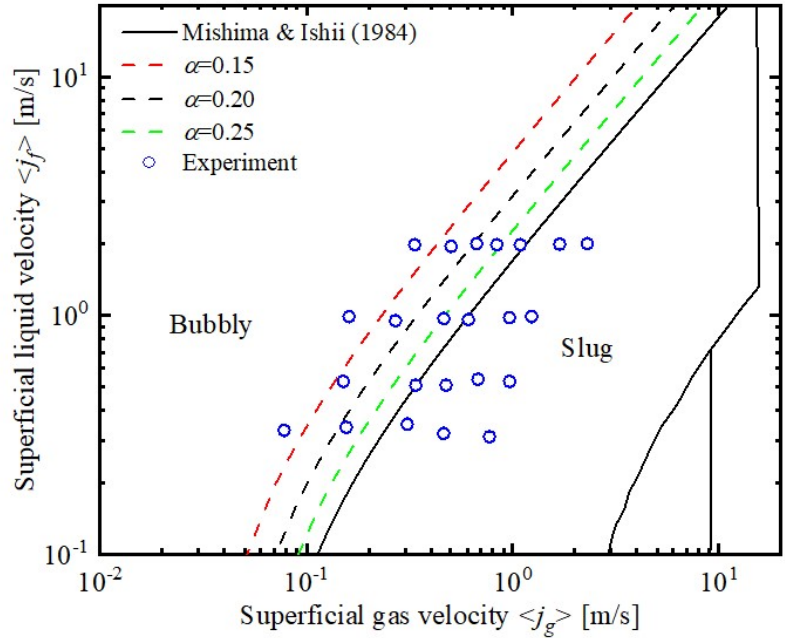

Figure 4: Test matrix.

## 2.3. Data verification

To verify the accuracy of the measurements, two methods are used in the present study. Firstly, the time-averaged local void fractions are converted into an average quantity by area-weighted averaging. The averaged value is compared with the impedance void meter measurement.[22] Secondly, the area-averaged superficial gas velocity is obtained using the local void fractions and the gas velocities, and it is compared with the superficial gas velocity measured using rotameters. As depicted in Fig. 5, good agreements are obtained between the void fraction measurements using conductivity probe and the impedance meter with an average uncertainty of 12.86%, and between the superficial gas velocity measurement using conductivity probe and rotameters with an average uncertainty of 13.06%.

Table 1: Flow conditions.

| $<j_f>$, m/s | $<j_{g,0}>$, m/s | | | | | | |
|---|---|---|---|---|---|---|---|
| Symbols | ■ | ● | ▲ | ▼ | ◆ | ◄ | ★ |
| 0.3 | 0.078 | 0.156 | 0.308 | 0.462 | 0.772 | | |
| $\alpha_{z/D=282}$ | 0.155 | 0.329 | 0.440 | 0.479 | 0.591 | | |
| **0.5** | **0.151** | **0.339** | **0.475** | **0.679** | **0.966** | | |
| $\alpha_{z/D=282}$ | 0.203 | 0.288 | 0.403 | 0.408 | 0.513 | | |
| **1.0** | **0.161** | **0.270** | **0.463** | **0.610** | **0.966** | **1.235** | |
| $\alpha_{z/D=282}$ | 0.126 | 0.225 | 0.253 | 0.346 | 0.477 | 0.660 | |
| **2.0** | **0.335** | **0.502** | **0.669** | **0.837** | **1.088** | **1.689** | **2.301** |
| $\alpha_{z/D=282}$ | 0.101 | 0.176 | 0.248 | 0.290 | 0.330 | 0.422 | 0.487 |

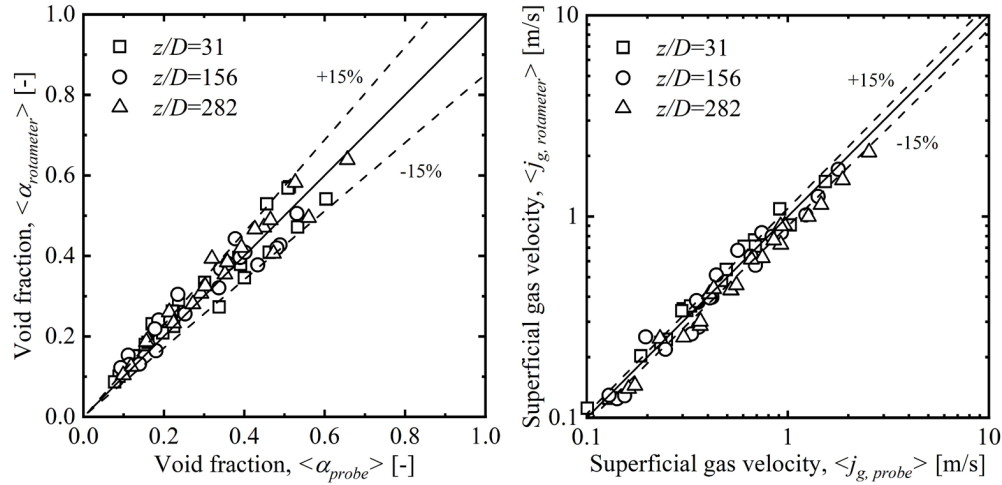

Figure 5: Verification of the four-sensor conductivity probe method with other calibration methods.

## 3. Results and discussion

### 3.1. Local measurement results

#### 3.1.1. Void fraction

Fig. 6 shows the void fraction profiles measured at $z/D = 282$. The profiles in each of the eight sub-figures are plotted together according to the superficial liquid velocities $<j_f>$ and bubble groups (Group-1 or Group-2). The sub-figures in the upper row of the figure are Group-1 void fraction profiles and those in the lower row are Group-2 void fraction profiles. The sub-figures in each column give the profiles with the same $<j_f>$. The symbols of each profile can be found in Table 1.

Overall, experimental data shows that unlike in bubbly flow, the void fraction distribution in the bubbly-to-slug transition flow has no wall peak distribution. All the experimental results in this study are center-peak distributed. This is because bubble size is usually large under these flow conditions. Due to the forces acting on the radial direction of the flow channel cross-section, such as shear-induced lift force and wall lubrication force, bubbles tend to migrate to the centerline of the flow channel. In comparison with the void distribution in relatively large diameter tubes (e.g. 25.4mm [23] or 50.8mm [10] ID tube), the wall peak distribution was observed under similar flow conditions. This difference of void distribution among different tube sizes is due to the relative size of the bubble to the flow channel size. With a smaller tube size, the relative bubble size is larger and there can be less freedom for the bubble to migrate in the radial direction of the flow channel. The distribution of the bubble can be less likely to show a peak near the wall. Another way to consider it is that with the same bubble size, bubbles can be closer to the centerline of a small flow channel than a large flow channel. Thus, the void distribution is hardly wall-peak distributed in a small diameter pipe. Another difference in terms of the local parameter profiles between the current study and studies in large diameter tubes is that the center part of profiles in the current study is slightly flat. This can be explained since the relative size of the bubble to the tube diameter is large in a small diameter tube, bubble that stays at the center of the flow channel can cover a larger area. This can also be seen more clearly from the bubble diameter profiles in Fig. 8.

The effect of the superficial liquid and gas velocities $<j_f>$ and $<j_g>$ on the void fraction profiles can be observed from the current experiments. The increase in $<j_f>$ can reduce void fraction and flatten the profile, while $<j_g>$ has the opposite effect. For the two-group void fraction distribution, with a constant $<j_f>$, the increase in superficial gas velocity $<j_g>$ increases the Group-2 void fraction with a center-peak distribution and suppresses the Group-1 void fraction. Under high superficial liquid and gas velocities ($<j_f> >= 1.0m/s$; $<j_g> >= 1.0m/s$), the Group-2 bubbles can push the Group-1 bubbles towards the wall, resulting in the Group-1 void fraction to be wall-peak distributed. This can be seen from Fig. 7c. In this situation when the Group-2 bubble appears to be dominant, the size of the Group-1 bubble tends to be small so that a wall-peak distribution can emerge. The effect of $<j_f>$ on Group-2 bubble structure can also be seen from the images, as high $<j_f>$ can make the interface of the Group-2 bubble becomes wavy. This is caused by the increase of the turbulence intensity.

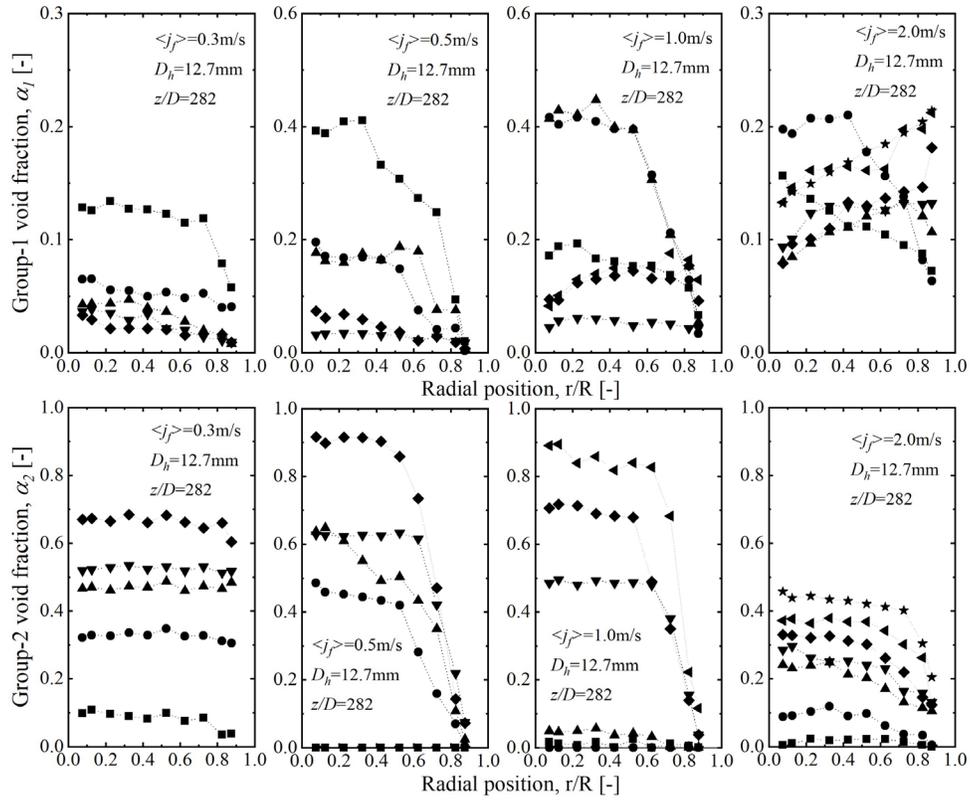

Figure 6: Local time-averaged void fraction profiles.

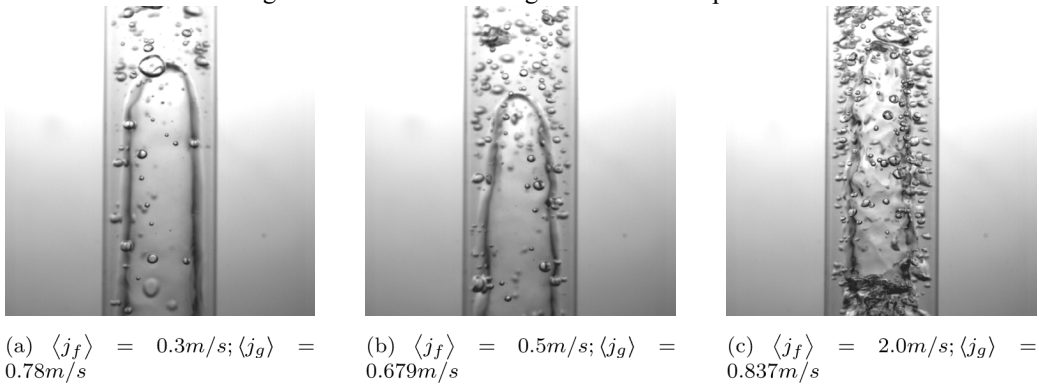

(a) $\langle j_f \rangle = 0.3 m/s; \langle j_g \rangle = 0.78 m/s$

(b) $\langle j_f \rangle = 0.5 m/s; \langle j_g \rangle = 0.679 m/s$

(c) $\langle j_f \rangle = 2.0 m/s; \langle j_g \rangle = 0.837 m/s$

Figure 7: Visual observations of slug bubble structure under different flow conditions taken at $z/D = 195$.

*3.1.2. Sauter mean diameter*

The bubble Sauter mean diameter profiles obtained from the current experimental study are presented in Fig. 8. Unlike in bubbly flow [24], the bubble Sauter mean diameter profile can have a center peak distribution in bubbly-to-slug flow transition flow. As discussed in the previous section, this is because large bubbles created by coalescence can migrate to the center of the flow channel due to the net forces in radial direction. Another observation from the results is that the Group-1 bubble size is reduced when the Group-2 bubble appears, as mentioned in the last section. Two reasons contribute to this phenomenon: firstly, large Group-1 bubbles coalesce and become Group2 bubbles, so the average Group-1 bubble size is reduced. Secondly, small bubbles are sheared off at the rim of large Group-2 bubbles. These bubbles tend to be smaller than 2mm, thus the average size of the Group-1 bubble decreases. Because of the second reason, the average Group-1 bubble size reduces more significantly under flow conditions with high superficial liquid velocities ($<j_f> >= 1.0$). The bubble shearing-off mechanism can be large under a high $<j_f>$ flow condition.

*3.1.3. Interfacial area concentration*

Fig. 9 shows the interfacial area concentration (IAC) profiles, which corresponds to those of void fraction profiles in Fig. 6. Similar to the previous experimental study, the Group-1 IAC profiles are similar to their corresponding void fraction profiles. However, the steepness of the IAC profile is different from the void fraction profile. This is because the bubble diameter varies in the radial direction of the flow channel cross-section. Fig. 8 gives the bubble Sauter mean diameter profiles, and it can be seen that Group-1 bubbles near the outer wall tend to be smaller. Group-2 IAC profile is wall-peak distributed under high superficial liquid velocity $<j_f> = 2.0 m/s$. As discussed in the previous section, the interface of the Group-2 bubble becomes wavy and high IAC values can be measured by the conductivity probe the edge of the large bubbles.

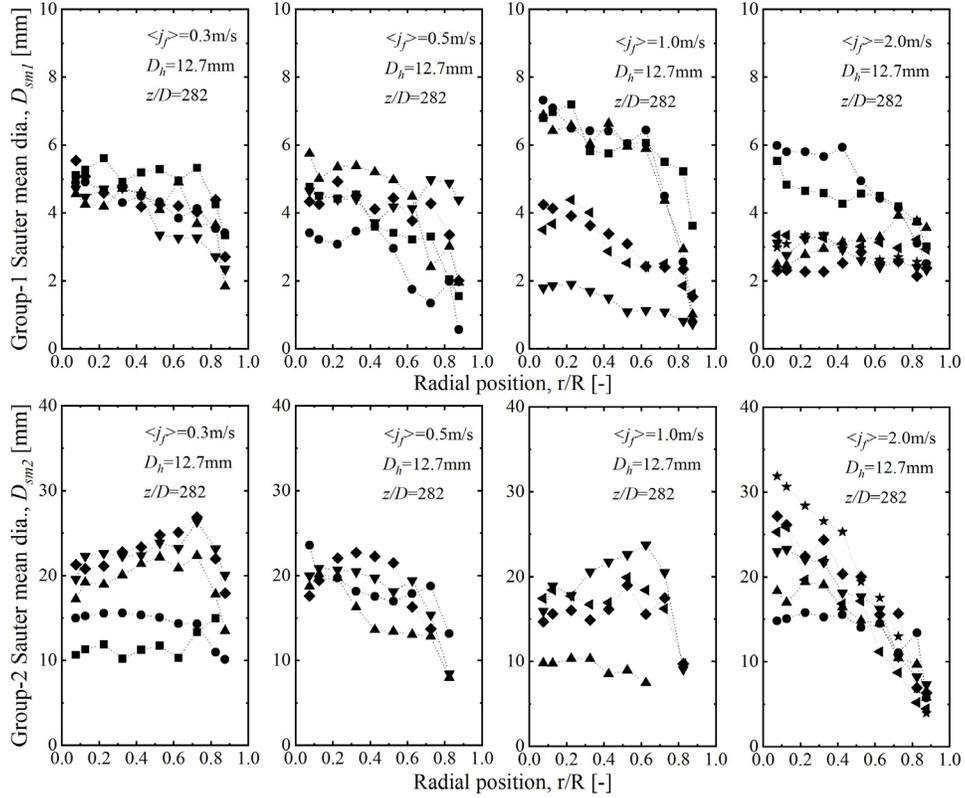

Figure 8: Local bubble Sauter mean diameter profiles.

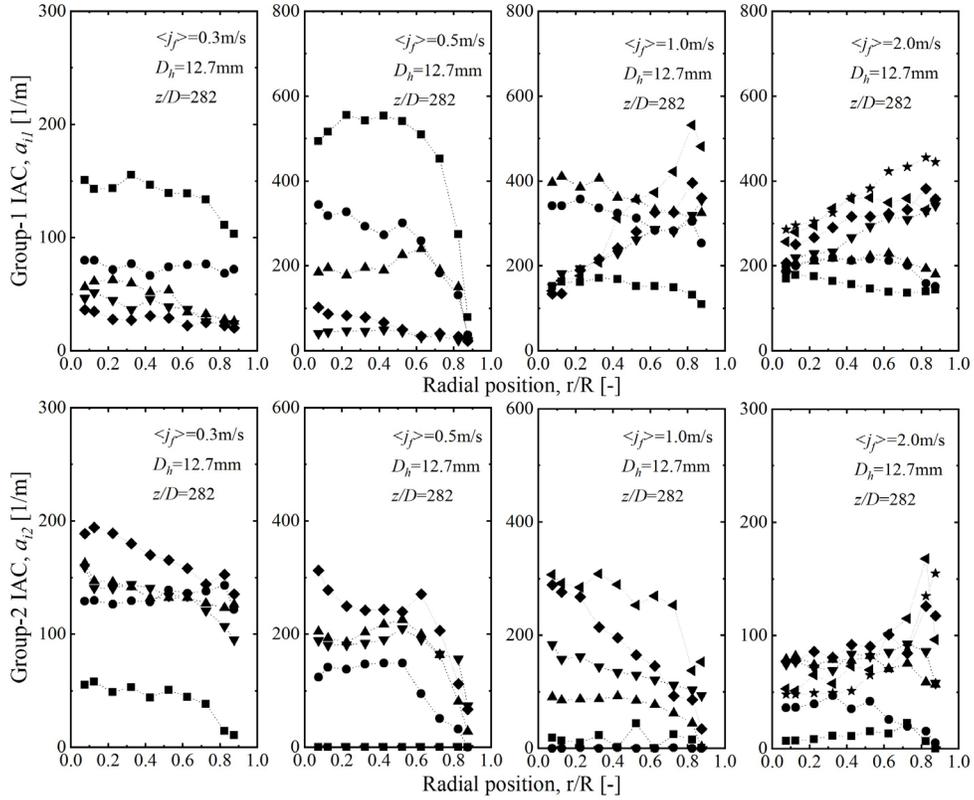

Figure 9: Local time-averaged interfacial area concentration profiles.

### 3.1.4. Interfacial velocity

Fig. 10 shows the interfacial velocity profiles, which correspond to those of void fraction profiles in Fig. 6. Similar to the experimental observations on previous studies [13, 24], for low superficial liquid velocities ($<j_f> \leq 0.5 m/s$), interfacial velocity profiles are usually flatter compared with the single-phase liquid velocity profile. The existence of bubbles in the liquid flow can flatten the liquid velocity profile. [24] For high $<j_f>$, the interfacial velocity profiles follow the power-law profile when flow is developed. The steepness of the interfacial velocity profile is related to both superficial liquid and gas velocities.

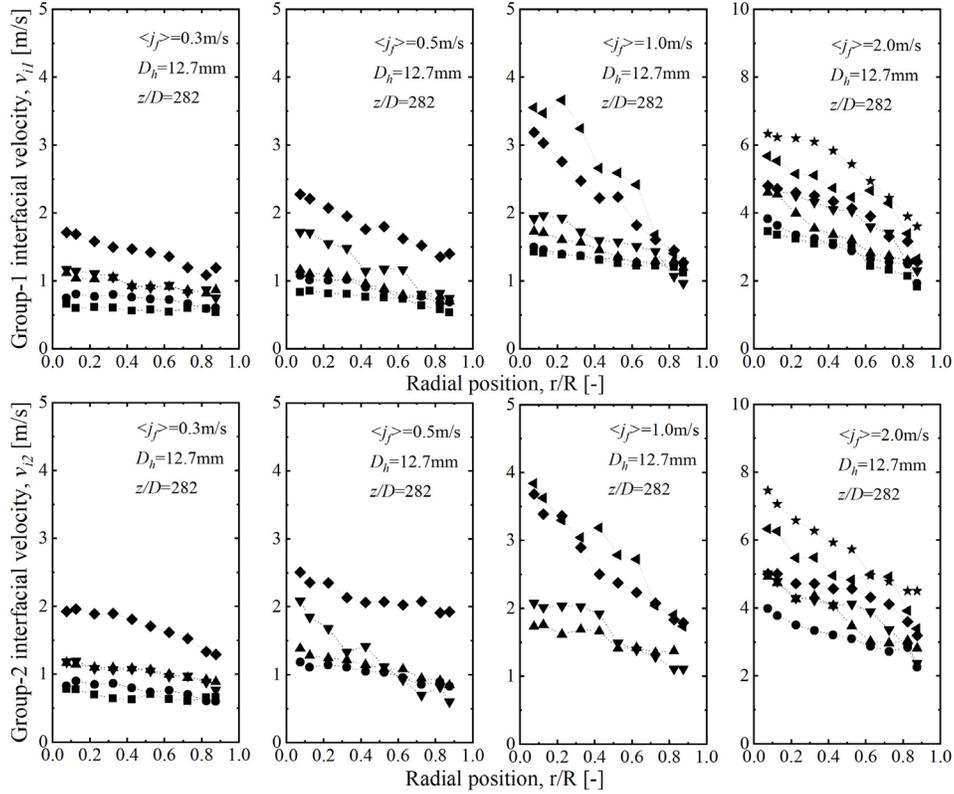

Figure 10: Local time-averaged interfacial velocity profiles.

*3.2. Correlation of distribution parameter*

From the previous discussion, the wall peak distribution of void fraction is not as significant as those in large diameter tube experiments. Similar observation was reported in the previous experimental study in a 9 mm ID tube [25]. The peak position of the void fraction profile of bubbly flow is closer to the centerline of the flow channel than observed in the 25.4 mm ID tube [26]. It indicates that the void fraction distribution profile is strongly related to the tube size. One easy way to understand this phenomenon is by considering bubbles with equal size in different sizes of tubes. Due to the small relative size of the bubble to the tube, bubbles in small diameter tubes are closer to the center of the tube.

In the Drift Flux model, the non-uniform distribution of two-phase flow is modeled by the distribution parameter, $C_0$. For subcooled boiling flow in small diameter round pipes, the correlation for distribution parameter is given by Ishii [21],

$$C_0 = (1.2 - 0.2\sqrt{\rho_g/\rho_f})\left(1 - e^{-18\langle\alpha\rangle}\right) \tag{2}$$

This correlation indicates that besides the properties of fluids, void fraction is the key parameter that affects the void distribution. Although this correlation was developed for subcooled boiling flows, it was also used for adiabatic air-water conditions. Noted that this correlation is applicable for flow conditions under $0 < \alpha < 0.25$, where the two-phase flow belongs to bubbly flow regime. Later on, Hibiki and Ishii [27] took into account the effect of the bubble size on the void distribution and formulated a new correlation,

$$C_0 = (1.2 - 0.2\sqrt{\rho_g/\rho_f})\left(1 - e^{-22(D_{sm})/D}\right) \tag{3}$$

Comparison of the performance between the two correlations was provided in the study of Hibiki and Ishii [27], which are given in Fig. 11. As can be seen from the figure, Ishii's model does not agree well with the experimental distribution parameters. The performance of Hibiki's model strongly depends on the tube size. From the two models, it should be noted that the void fraction is an experimental parameter describing the volume share of gas phase, namely a cubic-dimensional parameter, while the Sauter mean diameter describes the dimension of the average bubble size in a linear-dimensional space. Since the concept of distribution parameter is one-dimensional, and bubble distribution determined experimentally is determined using the onedimensional profiles, the averaged void fraction used in Ishii's model should be reformed to a linear-dimensional space. Thus, Ishii's model is modified
as,

$$C_0 = (1.2 - 0.2\sqrt{\rho_g/\rho_f})\left(1 - e^{-C\sqrt[3]{\langle\alpha\rangle}}\right) \tag{4}$$

Fig. 12 depicts the performance of the modified Ishii's correlation comparing with the data. Unlike the original Ishii's correlation, the modified correlation has better agreements with the experimental distribution parameters.

Another approach on improving modeling of the distribution parameter is to consider both void fraction and bubble size into account, since both of the two parameters have effects on the bubble distribution, as depicted in the conceptual schematics Fig. 13. Assuming that the two parameters affect the bubble distribution in the same order of magnitude, the correlation is formulated in the following form,

$$C_0 = (1.2 - 0.2\sqrt{\rho_g/\rho_f})\left(1 - e^{-C_1\langle D^*_{sm}\rangle - C_2\sqrt[3]{\langle\alpha\rangle}}\right) \tag{5}$$

where $\langle D_{sm}^* \rangle = \langle D_{sm} \rangle / D_h$ is the dimensionless Sauter mean diameter. $C_1$ and $C_2$ are coefficients determined experimentally. Fig. 14a and 14b show the comparison of the new-derived correlation and experimental distribution parameters from multiple studies. Based on these data, the coefficients are determined as $C_1 = 5$ and $C_2 = 3$. Table 2 shows the absolute relative errors of the distribution parameter correlations mentioned above against the data. It can be seen that both the modified Ishii's correlation and the new-derived correlation have better performance than the original Ishii's and Hibiki and Ishii's correlation.

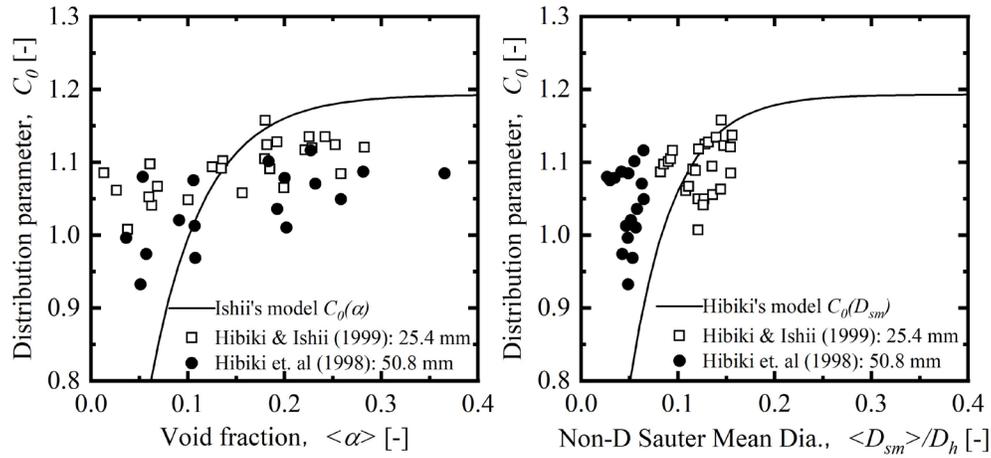

Figure 11: The comparison between the distribution parameter correlation, left): by Ishii (1977)[21] (Eq.2); right): by Hibiki and Ishii (2002)[27] (Eq.3) and the experimental distribution parameters [26, 28].[27]

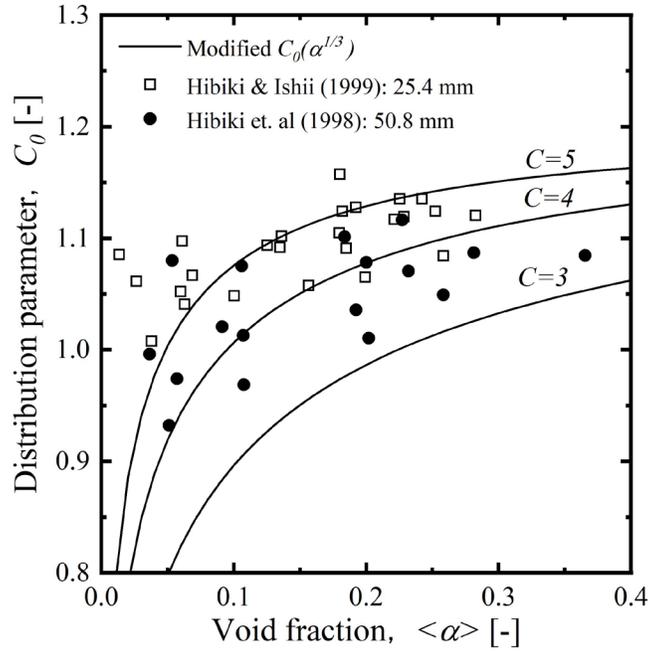

Figure 12: The comparison between the modified distribution parameter correlation based on Ishii (1977) (Eq.4) and the experimental distribution parameters [26, 28]

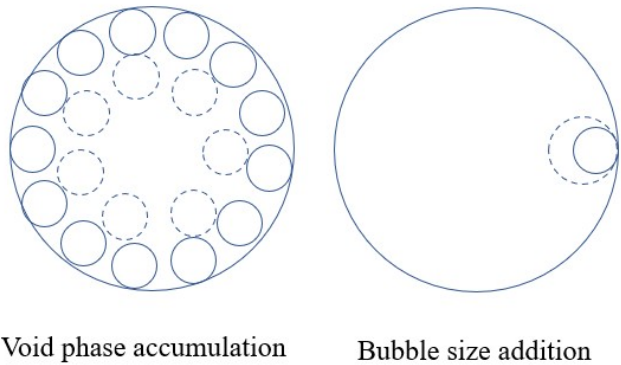

Figure 13: Schematic diagram of the bubble distribution modeling by separately considering the influential elements.

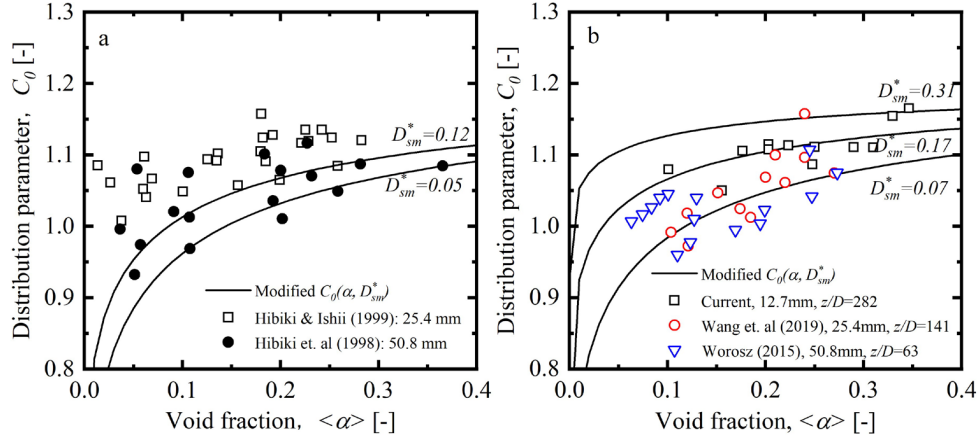

Figure 14: The comparison between new-derived distribution parameter correlation based on previous researches [21, 27] with the experimental distribution parameters in a): previous experimental studies [26, 28]; b): current and previous experimental studies [23, 10].

Table 2: Summary of absolute relative errors of the distribution parameter correlations against databases.

| Experiment | Ishii,1977[21] | Hibiki,2002[27] | Eq.4 | Eq.5 |
|---|---|---|---|---|
| Hibiki et al., 1999, 25.4mm[26] | 13.0% | 25.3% | 6.0% | 5.9% |
| Hibiki et al., 1998, 50.8mm[28] | 13.8% | 4.2% | 3.6% | 5.2% |
| Wang et.al, 2019, 25.4mm[23] | 7.5% | 10.0% | 6.9% | 4.9% |
| Worosz, 2015, 50.8mm[10] | 10.1% | 10.4% | 7.2% | 4.5% |
| Present, 12.7mm | 6.6% | 8.5% | 3.7% | 4.6% |
| Total | 10.8% | 11.2% | 5.3% | 5.0% |

*3.3. Averaged measurement results*

The averaged void fractions and IACs are obtained by integrating the time-averaged local measurement values over the flow channel. Fig. 15 shows the Group-1 and -2 area-averaged void fraction and IAC against superficial gas velocity $<j_g>$. Group-1 void fraction and IAC decrease, and Group-2 void fraction and IAC increase with the increase in $<j_g>$. As discussed before, Group-2 bubbles are formed by Group-1 bubbles coalescing. This transition between Group-1 and

-2 bubbles happens with a low $<j_g>$ when $<j_f>$ is low. For $<j_g>$ = 2.0$m/s$, Group-1 void fraction and IAC keep increasing when Group-2 bubble appears. As reported in the experimental study of 25.4mm ID tube [23], Group-1 void fraction and IAC still have a decreasing trend at similar conditions like in low $<j_g>$s. This indicates that the bubble coalescence rate in the 12.7mm ID tube is smaller than in the 25.4mm ID tube.

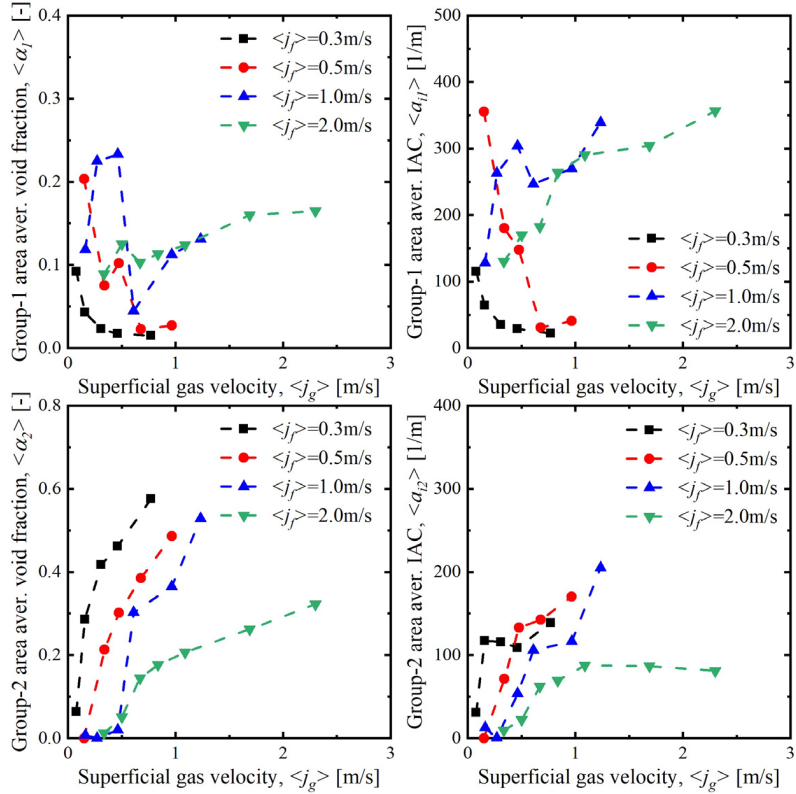

Figure 15: Area-averaged Group-1 and -2 void fractions and IACs.

## 4. Conclusions

This study provides local measurements of the void fraction, interfacial area concentration, and interfacial velocity in a 12.7 mm ID diameter tube using four-sensor electrical conductivity probe. 23 flow conditions in total focusing on the bubbly-to-slug transition flow regime are performed in this experiment. The accuracy of the measurement is cross-verified with impedance void meters and rotameters and the total average accuracy is within 15%. The characteristics of radial profiles of local parameters are discussed in this paper. The differences in terms of the local parameter distribution between the present study and the studies in larger sizes of tubes (25.4mm and 50.8mm) are observed. The void distribution is related to both void fraction and relative bubble size to the tube size. Therefore, a new correlation of distribution parameter is proposed based on the previous studies [21, 27]. The decrease of Group-1 void fraction and IAC with the increase of superficial gas velocity during the flow regime transition is observed in this study. The result also shows that the bubble coalescence rate is smaller in 12.7 mm ID tube than in 25.4 mm ID tube. This experimental study provides experimental references for the model development of flow regime transition and the Interfacial Area Transport Equation.